\documentclass[showpacs, showkeys, aps,preprint,twocolumn]{revtex4-1}%
\usepackage{amsfonts}
\usepackage{amsmath}
\usepackage{amssymb}
\usepackage{graphicx}
\usepackage{dcolumn}%
\providecommand{\U}[1]{\protect\rule{.1in}{.1in}}
\begin{document}
\title[Stochastic energy sink]{Stochastic energy sink in a low-dimensional Hamiltonian system}
\author{V.N. Pilipchuk}
\affiliation{Wayne State University \\ e-mail: pilipchuk@wayne.edu}
\eid{identifier}
\maketitle

%\date[Date text]{date}
%\received[Received text]{date}

%\preprint{HEP/123-qed}

%\author{Second Author}
%\affiliation{My Institution}
%\author{Third Author}
%\affiliation{Other Institution}

%\begin{abstract}
%Shell document for REV\TeX{} 4.

\textbf{Abstract} A few-degrees-of-freedom Hamiltonian model exhibiting
one-directional long-term trends in energy exchange flows is introduced. The
model includes a massive potential well - a container with one or few
relatively light non-interacting particles \ - attached to a linearly elastic
spring. No phenomenological dissipation is imposed, nevertheless, due to a
similarity of the container shapes to typical stochastic soft-wall billiards,
the energy is transferred from the container (donor) to the inner particles
(acceptor) in almost irreversible way during physically reasonable time
intervals. The potential well is introduced in such a way that, in the
rigid-body limit, it resembles either Sinai billiards or the so-called
Buminovich stadiums as the main geometrical parameter of the well switches its
sign. In particular, using the nonlinear normal mode stability concept reveals
conditions of stochasticity and determines the analogy with the dynamic
properties of billiards. Possible applications to the design of macro-level
energy absorbers, harvesters, and energy absorbing materials are discussed.

%\end{abstract}

%\volumeyear{year}
%\volumenumber{number}
%\issuenumber{number}

%\revised[Revised text]{date}

%\accepted[Accepted text]{date}

%\published[Published text]{date}

%\startpage{101}
%\endpage{102}
%\tableofcontents

\section{Introduction}

Modeling physical mechanisms of nonreciprocal energy exchange between coupled oscillatory subsystems represents a fundamental interdisciplinary problem. In particular, it may occur when designing molecular
structures with desired \textit{targeted energy transfer} properties
\cite{Aubry:2001}, \cite{Mackay:2003}, \cite{Maniadis:2005}, vibrational
energy harvesters \cite{Cottone:2009}, \cite{Gammaitoni:2009}, or mechanical
energy absorbers for controlling the structural dynamics
\cite{Vakakis:2009Springer}. In the latter case, the intent is to create a
relatively light device irreversibly absorbing the energy from the main structure. Since the Poincar\'{e} recurrence theorem prohibits such
effects to occur within the class of conservative systems, the irreversibility
is imposed phenomenologically using the conventional viscous damping. The core
of such studies is therefore analyses of the resonance dynamics with the goal to intensify the energy flow in a certain direction. The main problem is due
to the fact that directions of resonance energy flows depend upon initial
dynamic states. Therefore, stochasticity and ergodicity effects, allowing the
system to `forget' its initial states quickly enough, must be of interest in
this case. Furthermore, we show below that such dynamic properties generate
thermalization and dissipation, and thus irreversibility effects even when any
phenomenological dissipation is absent at all. \ Note that either direct
phenomenology or statistics of \textit{large systems} are usually involved for
modeling the dissipation \cite{Zwanzig:1973}, \cite{Jarzynski:1993}. The
present work however deals with the class of low-dimensional autonomous
Hamiltonian systems whose irregular stochastic dynamics are caused by
nonlinearities. A complete overview of this area, which has been under study
for quite a long time, is rather outside the scope of the present work. Let us
mention nevertheless that the corresponding phenomena may have very different
physical nature. For instance, nonlinearities dictating the geometry of
resonance manifolds, spectral overlaps, and modal interactions of the
resonance dynamics were expected to generate thermalization effects in the
Fermi--Pasta--Ulam (FPU) numerical test \cite{Zabusky:1965},
\cite{Zaslavskii:1972}. It was found though that the FPU multiple
degrees-of-freedom nonlinear chain model possesses certain properties of
integrable systems yet with a very complicated quasi-periodic behavior. An
alternative class of models emerged from the theory of billiards
\cite{Sinai:1970}, \cite{Buminovich:1985}. The billiard nonsmooth dynamics are
quite opposite to the resonance quasi harmonic dynamics and therefore usually
described in a different way through the discrete time mapping. Although
trajectories of particles may be geometrically simple inside billiard domains,
conditioning collisions of particles with billiard walls may appear to be
quite complicated \cite{RyabovLoskutov2010}.

\begin{figure}[ptb]%
\centering
\includegraphics[
height=5.6936cm,
width=7.5978cm
]%
{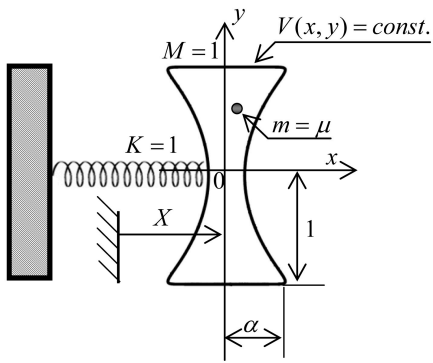}%
\caption{A massive potential container oscillating with one small particle
inside ($k=1)$, which is interacting with potential walls of different
`stiffness' and contour shapes; see Fig. 2.}%
\label{fig1}%
\end{figure}

\begin{figure}[ptb]%
\centering
\includegraphics[
height=7.6273cm,
width=7.5978cm
]%
{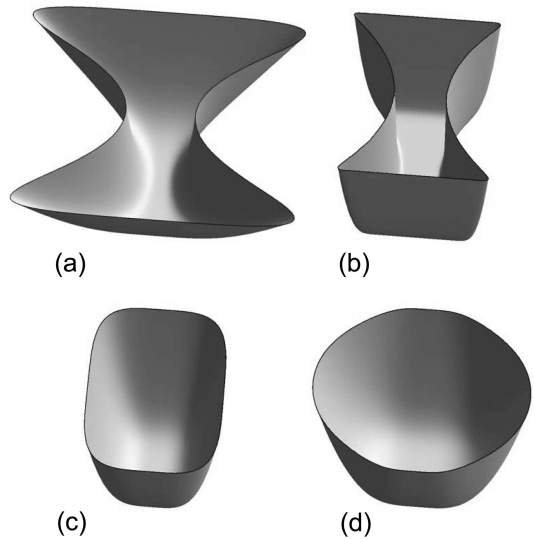}%
\caption{Different shapes of the potential container obtained for $\alpha=1/2$
and $\gamma=1.0$, and the following wall's stiffness ($n$) and contour's
curvature ($\beta$) parameters: a) $n=2$, $\beta=0.3$; b) $n=10$, $\beta=0.3$;
c) $n=2$, $\beta=0.0$; and d) $n=2$, $\beta=-0.3$.}%
\label{fig2}%
\end{figure}

The present approach is two-fold. On one hand, we essentially employ the
analogy with classical billiards for physical interpretations as soon as the
soft-wall potential container of the present model can degenerate into two
different basic types of billiards in the rigid-body asymptotic limit. On the
other hand, we apply some analytical and numerical tools of smooth dynamics in
order to determine conditions of stochasticity. The goal is to illustrate the
possibility of a few degrees-of-freedom `energy sink' within the class of
Hamiltonian systems. Note that Poincar\'{e} recurrence is theoretically
inevitable in such case, however, we show that its time span extends far
beyond any natural time intervals of the system. In our case\textit{, the
effect is due to specific shapes of the soft-wall container}, represented by
the moving potential well; see Figs.\ref{fig1} and \ref{fig2}. Although smooth
potentials are shown, the physical justification of the results can be found
within the theory of typical rigid-wall billiards, where the thermodynamic
properties are associated with ergodicity and positive Lyapunov exponents due
to the scattering effect of boundaries \cite{Sinai:1970}. Few decades ago it
was also found that the presence of convex scatters is not necessary for
generating the dynamic stochasticity. The corresponding billiard shapes
resemble stadium fields and are known as Buminovich stadia
\cite{Buminovich:1985}. For that reason, the present model is developed in
such a way that, in the rigid-body limit, it can degenerate into one or
another type of billiards when changing the sign of the main parameter of the
potential well, $\beta$; see expression (\ref{V definition}) below.
%TCIMACRO{\FRAME{ftbpFU}{7.5978cm}{5.6936cm}{0pt}{\Qcb{A massive potential
%container oscillating with one small particle inside ($k=1)$, which is
%interacting with potential walls of different `stiffness' and contour shapes;
%see Fig. 2.}}{\Qlb{fig1}}{figure1.eps}{\special{ language "Scientific Word";
%type "GRAPHIC";  maintain-aspect-ratio TRUE;  display "USEDEF";
%valid_file "F";  width 7.5978cm;  height 5.6936cm;  depth 0pt;
%original-width 3.1531in;  original-height 2.3557in;  cropleft "0";
%filename 'Figure1.eps';file-properties "XNPEU";}} }%
%BeginExpansion
%EndExpansion
%TCIMACRO{\FRAME{ftbpFU}{7.5978cm}{7.6273cm}{0pt}{\Qcb{Different shapes of the
%potential container obtained for $\alpha=1/2$ and $\gamma=1.0$, and the
%following wall's stiffness ($n$) and contour's curvature ($\beta$) parameters:
%a) $n=2$, $\beta=0.3$; b) $n=10$, $\beta=0.3$; c) $n=2$, $\beta=0.0$; and d)
%$n=2$, $\beta=-0.3$.}}{\Qlb{fig2}}{figure2.eps}%
%width 7.5978cm;  height 7.6273cm;  depth 0pt;  original-width 4.273in;
%original-height 4.2886in;  cropleft "0";  croptop "1";  cropright "1";
%cropbottom "0";  filename 'Figure2.eps';file-properties "XNPEU";}} }%
%BeginExpansion
%EndExpansion
%TCIMACRO{\FRAME{ftbpFU}{7.5978cm}{10.5037cm}{0pt}{\Qcb{Bifurcation diagrams
%revealing irregular stochastic motions in different negative and positive
%intervals of the curvature parameter $\beta$ for a) $n=2$ - softer walls and
%b) $n=10$ - stiffer wall containers; $\alpha=1/2$, $\gamma=1.0$.}}{\Qlb{fig3}%
%}{figure3.eps}{\special{ language "Scientific Word";  type "GRAPHIC";
%width 7.5978cm;  height 10.5037cm;  depth 0pt;  original-width 3.0424in;
%original-height 4.216in;  cropleft "0";  croptop "1";  cropright "1";
%cropbottom "0";  filename 'Figure3.eps';file-properties "XNPEU";}} }%
%BeginExpansion
\begin{figure}[ptb]%
\centering
\includegraphics[
height=10.5037cm,
width=7.5978cm
]%
{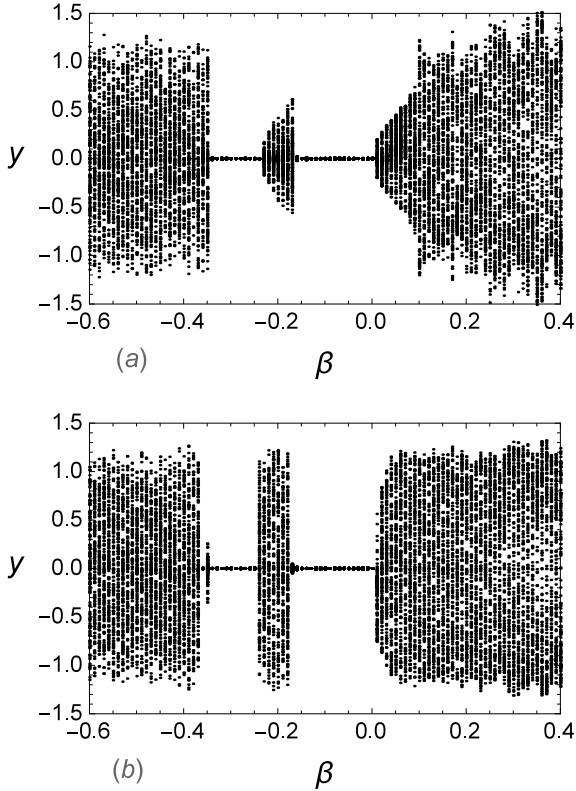}%
\caption{Bifurcation diagrams revealing irregular stochastic motions in
different negative and positive intervals of the curvature parameter $\beta$
for a) $n=2$ - softer walls and b) $n=10$ - stiffer wall containers;
$\alpha=1/2$, $\gamma=1.0$.}%
\label{fig3}%
\end{figure}
%EndExpansion
Finally, the main specific of the present modeling is that the soft-wall
potential container possesses a finite mass attached to a linearly elastic
spring and thus can dynamically interact with the contained particles creating
two-way energy exchange flows. Nevertheless, the main trend of the energy
exchange is shown to be one-directional. A brief physical explanation is that
vibrations of the massive potential container are quasi regular, while the
dynamics of relatively light particle(s) inside the container are quasi
stochastic. Also, while the container's motion is one-dimensional, the
particle(s) path covers two-dimensional domains. Such specifics create
different mechanisms for different directions of the energy exchange in such a
way that one of the directions becomes dominant.

\section{Model description}

The model is illustrated in Fig.\ref{fig1}, where a massive harmonic
oscillator $M$ represents a two-dimensional container described by the
potential energy $V=V(x,y)$. The container is shown by a typical level line of
the potential energy in the non-inertial Cartesian frame $xy$ associated with
the axes of symmetry of the potential well. The container includes $k$
relatively light non-interacting particles of the total mass $m<<M$, which is
driven by its interactions with the potential walls as the container is given
some initial energy. The total mass of particles is fixed while their number
can be different so that $\mu=m/k$ is a mass of each particle. A practical
macro-level design for such a model may include $k$ parallel containers in
order to exclude interactions between the particles. As mentioned in
Introduction, we intentionally assume no phenomenological dissipation in the
system, in other words, the total energy of the oscillator with particles is
conserved, and thus the system is Hamiltonian. Therefore, elastic collisions
with the container's walls affect also the dynamics of container however in a
less dramatic way. The presence of such a feedback, which is usually ignored
in statistical studies, represents a key assumption of the present work whose
purpose is the analysis of recurrence effects versus contour shapes of the
potential energy. Another difference with the typical statistical studies of
non-interacting gas models is that the number of particles, $1\leq$ $k\leq5$,
is rather insufficient to provide the statistics of large numbers. Assuming no
gravity is present, the potential energy is modeled with the function%
\begin{equation}
V=\frac{\gamma}{2n}\left[  \left(  \frac{x}{\alpha+\beta(y^{2}-1)}\right)
^{2n}+y^{2n}\right]  \label{V definition}%
\end{equation}

As follows from Fig.\ref{fig2}, the phenomenological expression
(\ref{V definition}) provides a convenient way to modeling qualitatively
different shapes of two-dimensional containers with `soft walls,' where
$\beta$ is the main geometrical parameter linked to the contour's curvature
as
\begin{equation}
\kappa=\frac{2\beta}{(1+4\beta^{2}y^{2})^{3/2}} \label{kappa}%
\end{equation}
for $x>0$, $-1<y<1$, as $n\longrightarrow\infty$.

Therefore $\beta$ is simply one half of the curvature at $y=0$. \ As mentioned
in Introduction, the model (\ref{V definition}) enables us to analyze soft
analogs of two different types of stiff boundaries. Namely, when $\beta>0$, we
have a soft model of billiards with scattering boundaries \cite{Sinai:1970},
while the case $\beta<0$ gives a soft approximation for the so-called
Buminovich stadiums \cite{Buminovich:1985}. In our model, $n$ is a wall
stiffness parameter, such that walls become asymptotically stiff as
$n\longrightarrow\infty$. Note that the case of soft walls seems to be more
realistic on physical view. Also, soft walls are more convenient from the
standpoint of simulations since no conditioning is required for collisions
against the walls. Now the model dynamics can be described by the Lagrangian%
\begin{equation}
L=\frac{1}{2}\dot{X}^{2}-\frac{1}{2}X^{2} \label{Lagr}%
\end{equation}%
\[
+\sum\limits_{j=1}^{k}\left\{  \frac{\mu}{2}\left[  \left(  \dot{X}+\dot
{x}_{j}\right)  ^{2}+\dot{y}_{j}^{2}\right]  -V_{j}\right\}
\]
where $V_{j}=V(x_{j},y_{j})$ is the potential energy of the \textit{j}th
particle, and overdots mean differentiation with respect to time $t$.

The corresponding Hamiltonian is obtained via the Legendre transform
\begin{equation}
H=P\dot{X}+\sum\limits_{j=1}^{k}p_{j}\dot{x}_{j}-L \label{Hamiltonian}%
\end{equation}
where $P=\partial L/\partial\dot{X}$ and $p_{j}=\partial L/\partial\dot{x}%
_{j}$ are linear momenta.
%TCIMACRO{\FRAME{ftbpFU}{7.5981cm}{11.4334cm}{0pt}{\Qcb{Trajectories of the
%particle inside the potential container in the short-term interval $0\leq
%t/T\leq100$ at $\alpha=1/2$, $\gamma=1.0$, $n=10$, and different contour
%shapes: a) $\beta=-0.5$, b) $\beta=-0.3$, c) $\beta=-0.2$, d) $\beta=-0.14$,
%e) $\beta=0.009$, and f) $\beta=0.3$; in all cases, the initial position of
%particle is ($x$,$y$) $=$($0.0$,$0.01$) with zero velocity.}}{\Qlb{fig4}%
%}{figure4.eps}{\special{ language "Scientific Word";  type "GRAPHIC";
%width 7.5981cm;  height 11.4334cm;  depth 0pt;  original-width 2.8314in;
%original-height 4.273in;  cropleft "0";  croptop "1";  cropright "1";
%cropbottom "0";  filename 'Figure4.eps';file-properties "XNPEU";}} }%
%BeginExpansion
\begin{figure}[ptb]%
\centering
\includegraphics[
height=11.4334cm,
width=7.5981cm
]%
{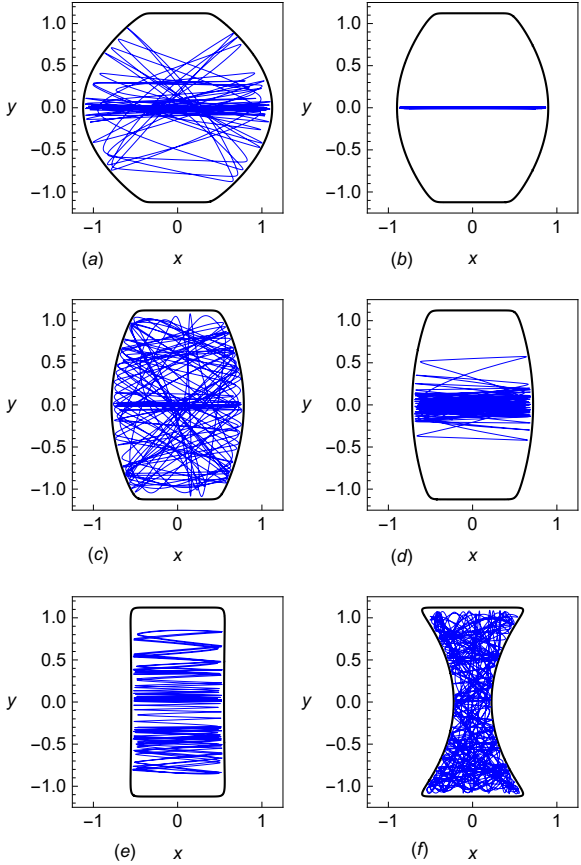}%
\caption{Trajectories of the particle inside the potential container in the
short-term interval $0\leq t/T\leq100$ at $\alpha=1/2$, $\gamma=1.0$, $n=10$,
and different contour shapes: a) $\beta=-0.5$, b) $\beta=-0.3$, c)
$\beta=-0.2$, d) $\beta=-0.14$, e) $\beta=0.009$, and f) $\beta=0.3$; in all
cases, the initial position of particle is ($x$,$y$) $=$($0.0$,$0.01$) with
zero velocity.}%
\label{fig4}%
\end{figure}
%EndExpansion
%TCIMACRO{\FRAME{ftbpFU}{7.5981cm}{7.7211cm}{0pt}{\Qcb{Time histories of the
%oscillator's energy corresponding to different contour shapes in Fig. 4;
%$\left\langle E\right\rangle $ is a mean value over the ensemble of 50 runs
%with random initial positions of the particle in the domain
%\{$-0.01<x(0)<0.01$, $-0.01<y(0)<0.01$\}; the parameters are $n=10$, $k=1$,
%$\alpha=1/2$.}}{\Qlb{fig5}}{figure5.eps}%
%width 7.5981cm;  height 7.7211cm;  depth 0pt;  original-width 5.2338in;
%original-height 5.3186in;  cropleft "0";  croptop "1";  cropright "1";
%cropbottom "0";  filename 'Figure5.eps';file-properties "XNPEU";}} }%
%BeginExpansion
\begin{figure}[ptb]%
\centering
\includegraphics[
height=7.7211cm,
width=7.5981cm
]%
{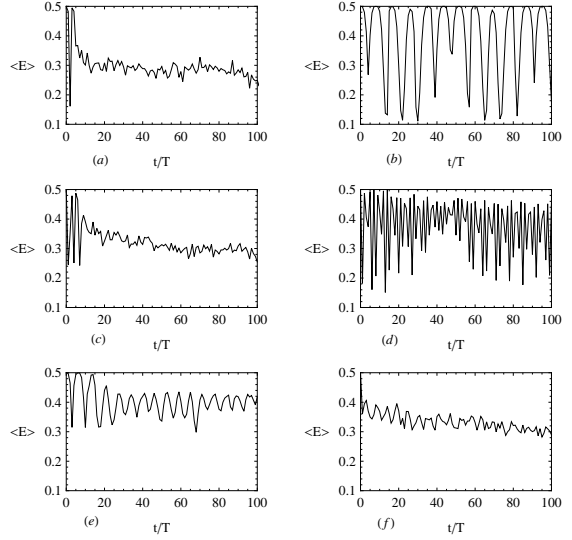}%
\caption{Time histories of the oscillator's energy corresponding to different
contour shapes in Fig. 4; $\left\langle E\right\rangle $ is a mean value over
the ensemble of 50 runs with random initial positions of the particle in the
domain \{$-0.01<x(0)<0.01$, $-0.01<y(0)<0.01$\}; the parameters are $n=10$,
$k=1$, $\alpha=1/2$.}%
\label{fig5}%
\end{figure}
%EndExpansion

\section{Analogies with billiards}

As follows from bifurcation diagrams obtained for the case of a single
particle, $k=1$, the soft wall dynamics are rather dictated by the
corresponding billiard shape in rigid-body limit; see Fig.\ref{fig3}.
\ Namely, comparing the cases of softer ($n=2$) and stiffer ($n=5$) walls,
illustrated by the fragments (a) and (b), respectively, shows qualitatively
similar dynamic behaviors in similar intervals of the parameter $\beta$. The
analogy with billiard dynamics is supported by different shapes of
trajectories, obtained even with stiffer walls ($n=10$); see Figs.\ref{fig4}
and \ref{fig5}. As seen from Fig.\ref{fig4}, container contours and the
corresponding trajectories of the particle can take quite different shapes
through the interval $-0.5<\beta<0.4$. In particular, the particle path is
either quite regular or chaotic depends upon the number $\beta$; compare to
the bifurcation diagrams shown in Fig.\ref{fig3}. When contours possess
convexities towards the inner domain ($\beta>0$), the presence of
irregularities comes of no surprise due to instabilities caused by the
scattering effects. The case $\beta<0$ appears to be more complicated since
both chaotic and regular motions are possible in different intervals of the
curvature. A geometrical nature of such phenomena is likely similar to that
was revealed in the case of Buminovich stadia \cite{Buminovich:1985}, in which
the exponential divergence of orbits is possible without convex scatters.

As follows from the above remarks, a possible approach to explanation of
regular and chaotic behaviors of the particle may employ transition to the
rigid-body billiard limit $n\rightarrow\infty$. In this case, the analysis can
be conducted in a purely geometrical however quite complicated way including
conditioning dictated by the billiard's stiff boundaries. Alternatively, we
use the nonlinear normal modes (NNMs) stability concept
\cite{Vakakis:1996Wiley} with the analytical tool of Floquet theory.

\section{NNM's Instability and Stochasticity}

In the case of a single particle, $k=1$, there is a NNM whose trajectory is
described by the equation $y_{1}=0$ as dictated by the symmetry of potential
well. Although such a trajectory can be only observed under specific initial
conditions, its local stability properties appear to have a global effect on
the dynamics inside the well. In order to confirm such an observation, let us
linearize the model with respect to the coordinate $y_{1}$ and re-scale the
coordinates and time as%
\begin{equation}
\{\bar{x},\bar{y},\bar{X},\bar{t}\}=\frac{1}{\alpha-\beta}\left\{
x,y_{1},X,t\sqrt{\frac{\gamma}{\mu}}\right\}  \label{rescale}%
\end{equation}

As follows from Fig.\ref{fig1} and definition (\ref{V definition}), the number
$\alpha-\beta$ becomes the least distance from the origin to a `vertical'
potential wall along the horizontal axis in the rigid-body limit
$n\rightarrow\infty$. Now let us skip the overbars in notations (\ref{rescale}%
) and consider the differential equations of motion of the particle inside the
potential container
\begin{align}
\ddot{x}+x^{2n-1}  &  =\frac{\mu}{1+\mu}\left[  \ddot{x}+\frac{(\alpha
-\beta)^{2}}{\gamma}X\right] \label{x-oscill}\\
\ddot{y}+\lambda x^{2n}y  &  =0 \label{y-oscill}%
\end{align}
where $\lambda=-2\beta(\alpha-\beta)$.

Using the assumption $\mu\ll1$ and ignoring the right-hand side of equation
(\ref{x-oscill}) gives an unloaded essentially nonlinear oscillator with the
power-form characteristic. Such oscillators were considered by Lyapunov when
analyzing the degenerated case of his stability of motions theory
\cite{Liapunov:2000Stability}. The period is calculated through the well
tabulated gamma-function $\Gamma$ as%
\begin{align}
T_{n}  &  =\frac{4\sqrt{n}}{A_{0}^{n-1}}\int\limits_{0}^{1}\frac{dz}%
{\sqrt{1-z^{2n}}}\label{T}\\
&  =\frac{2}{A_{0}^{n-1}}\sqrt{\frac{\pi}{n}}\frac{\Gamma(1/(2n))}%
{\Gamma((1+n)/(2n))}\nonumber
\end{align}
where $A_{0}=(2nE_{n})^{1/(2n)}$ is the amplitude versus the total energy
$E_{n}=\dot{x}^{2}/2+x^{2n}/(2n)$.%

%TCIMACRO{\FRAME{ftbpFU}{7.6552cm}{5.5553cm}{0pt}{\Qcb{Vibro-impact oscillator
%and its dynamic state functions; the relationship $\dot{\tau}^{2}=1$ holds
%almost everywhere, therefore $\dot{\tau}$ can be viewed as a unipotent of the
%so-called hyperbolic (Clifford's) algebra.}}{\Qlb{fig6}}{figure6.eps}%
%width 7.6552cm;  height 5.5553cm;  depth 0pt;  original-width 2.9698in;
%original-height 2.1473in;  cropleft "0";  croptop "1";  cropright "1";
%cropbottom "0";  filename 'Figure6.eps';file-properties "XNPEU";}} }%
%BeginExpansion
\begin{figure}[ptb]%
\centering
\includegraphics[
height=5.5553cm,
width=7.6552cm
]%
{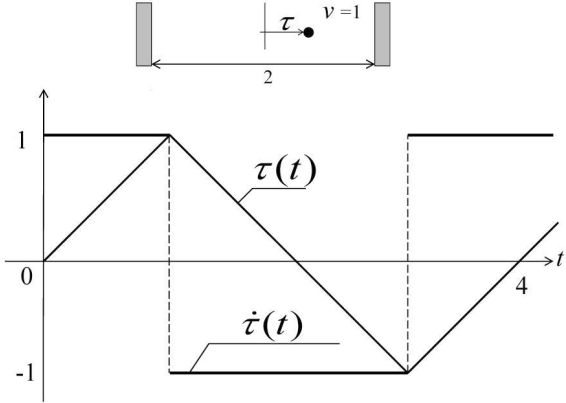}%
\caption{Vibro-impact oscillator and its dynamic state functions; the
relationship $\dot{\tau}^{2}=1$ holds almost everywhere, therefore $\dot{\tau
}$ can be viewed as a unipotent of the so-called hyperbolic (Clifford's)
algebra.}%
\label{fig6}%
\end{figure}
%EndExpansion

Although different versions of special functions have been suggested to
describe the temporal shape of vibrations $x(t)$, we prefer to use an
approximate solution in terms of elementary functions incorporating the
asymptotic of large exponents with a reasonable, for the present case,
precision. Such an approximation is obtained in the form of power series
expansion with respect to the periodic triangle wave function%
\begin{equation}
\tau=\tau\left(  \frac{t}{T_{n}}\right)  =\frac{2}{\pi}\arcsin\sin\frac{2\pi
t}{T_{n}}\text{,\qquad}|\tau|\leq1 \label{tau intro}%
\end{equation}

A physical interpretation of function (\ref{tau intro}) is explained in
Fig.\ref{fig6}. Namely, we use the dynamic states of elementary vibro-impact
oscillator as a basis for the description of non-smooth as well as smooth
vibrations based on the fact that the set $\{1,\dot{\tau}\}$ generates the
so-called hyperbolic algebraic structures \cite{Pilipchuk:2010Springer}. A
starting point of the corresponding analytical procedure is the periodic
temporal substitution $t\rightarrow\tau$. This leads to a boundary value
problem in the standard interval, $-1\leq\tau\leq1$, which is solved
iteratively to give $T_{n}$-periodic solution%
\begin{equation}
x=U(\tau)=A\left(  \tau-\frac{\tau^{2n+1}}{2n+1}-R_{2}-\cdots\right)
\label{X example}%
\end{equation}
where the constant parameter $A$ and the amplitude $A_{0}$ are coupled by the
equation $A_{0}=U(1)$, and the following quantities provide estimates for
high-order terms of the successive approximations:%
\[
R_{2}(\tau,n)=\frac{1}{2}\frac{2n-1}{2n+1}\left(  \frac{\tau^{2n+1}}%
{2n+1}-\frac{\tau^{4n+1}}{4n+1}\right)
\]%
\begin{equation}
0<R_{i}<\frac{(2n-1)\left\vert \tau\right\vert ^{2n+1}}{2^{i-1}(2n+1)^{2}}
\label{estimates}%
\end{equation}

In particular, expressions (\ref{estimates}) indicate that series
(\ref{X example}) converge quite slowly. However, the asymptotic of large
exponents essentially improves precision of the truncated series even though
first few terms of the series are included. Obviously, as $n\rightarrow\infty
$, the oscillator becomes a particle between two perfectly stiff barriers
$x=\pm1$; see the upper fragment in Fig. \ref{fig6}.

\begin{figure}[ptb]%
\centering
\includegraphics[
height=5.3532cm,
width=7.5981cm
]%
{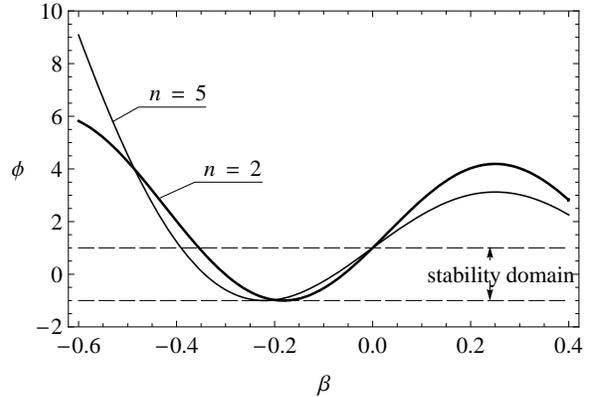}%
\caption{The stability domain.}%
\label{fig7}%
\end{figure}

Due to periodicity of solution (\ref{X example}), equation (\ref{y-oscill})
represents Hill's equation. Note that, compared to the solution $x(t)$, the
period of the coefficient is reduced as many as twice due to the even exponent
$2n$. Nonetheless, the period $T_{n}$ still leads to the same stability
conditions determined by Floquet multipliers $\rho_{1,2}=\phi\pm\sqrt{\phi
^{2}-1}$. The number $\phi=[y_{1}(T_{n})+\dot{y}_{2}(T_{n})]/2$ is calculated
through the two fundamental solutions of equation (\ref{y-oscill}), $y_{1}(t)$
and $y_{2}(t)$, such that $y_{1}(0)=1$,\quad$\dot{y}_{1}(0)=0$ and
$y_{2}(0)=0$,\quad$\dot{y}_{2}(0)=1$, respectively. Based on the number $\phi
$, the solution $y(t)$ is unstable if $\phi^{2}>1$, and stable if $\phi^{2}%
<1$. If $\phi^{2}=1$, there exist a periodic solution of equation
(\ref{y-oscill}). Fig.\ref{fig7} illustrates the dependence $\phi=\phi(\beta
)$, which actually shows how the curvature of potential contour (\ref{kappa})
affects stability of the NNM $y=0$. In particular, stability domains appear to
be in a reasonable compliance with those captured with bifurcation diagrams in
Fig.\ref{fig3}(a), and (b). The relatively narrow instability regions inside
the stability intervals are rather caused by the boundary $\phi=-1$ touched by
the curves $\phi=\phi(\beta)$; compare Fig.\ref{fig3} to Fig.\ref{fig7}. Note
that the bifurcation diagrams impose no conditions on the coordinate $y$,
whereas the Floquet stability analysis is justified locally, near the axis of
symmetry $y=0$. In addition, ignoring the right-hand side of equation
(\ref{x-oscill}) is equivalent to a fixed potential container, whereas the
bifurcation diagrams in Fig.\ref{fig3} were obtained with the oscillating
container. Since both results are nevertheless in match, we can expect that
the local stability properties of the NNM $y=0$ determine qualitative features
of the global dynamics of entire system (\ref{Lagr}).%

%TCIMACRO{\FRAME{ftbpFU}{7.5981cm}{5.3532cm}{0pt}{\Qcb{The stability domain.}%
%}{\Qlb{fig7}}{figure7.eps}{\special{ language "Scientific Word";
%type "GRAPHIC";  maintain-aspect-ratio TRUE;  display "USEDEF";
%valid_file "F";  width 7.5981cm;  height 5.3532cm;  depth 0pt;
%original-width 4.1632in;  original-height 2.9222in;  cropleft "0";
%filename 'Figure7.eps';file-properties "XNPEU";}} }%
%BeginExpansion
%EndExpansion

\section{Recurrence and dissipative effects}

Assuming that the main oscillator of mass $M$ is given some initial energy,
while the inner particles are in rest, we analyze the process of energy
transfer from the oscillator to the particles at different magnitudes of the
parameter $\beta$. Since the model (\ref{Lagr}) is conservative then both the
direct and reversed energy flows are possible. However, the energy exchange
dynamics look different in different intervals of the parameter $\beta$ as
confirmed already by Figs.\ref{fig4} and \ref{fig5}, where different fragments
of Fig.\ref{fig5} relate to the same cases in Fig.\ref{fig4}. In order to
eliminate the possible influence of initial conditions, we consider the mean
energy flow over ensembles of randomly taken initial positions of the particle
in a narrow area near zero. It is seen that the trend of the energy outflow
from the donor is associated with the stochastic behavior of the receiver, and
it may take place in both negative and positive unstable sub intervals of the
parameter $\beta$.

\begin{figure}[ptb]%
\centering
\includegraphics[
height=5.0676cm,
width=7.5981cm
]%
{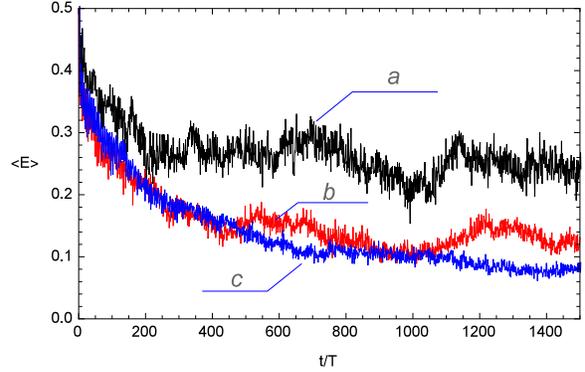}%
\caption{Container's energy during the temporal interval $0\leq t/T\leq1500$
in the case of convex scatters ( $\beta=0.3$ ) with the following numbers of
particles: a) $k=1$, b) $k=3$, and c) $k=5$; the averaging is taken over
ensemble of 25 runs over random initial positions of particles in the domain
\{$-0.01<x(0)<0.01$, $-0.01<y(0)<0.01$\}.}%
\label{fig8}%
\end{figure}

Below we analyze long-term trends of the energy flow by increasing the
observation interval to 1500 eigen periods of the oscillator, $0\leq
t/T\leq1500$. Consider first the case of scattering convex potential,
$\beta=0.3$. The result of simulations, which is represented in Fig.\ref{fig8}%
, confirms the presence of energy outflow effect at different however still
small number non-interacting particles inside the potential container. Recall
that the total mass of the energy absorbing particles remains fixed as
$m=k\mu=0.2$. However, it is seen that the effect becomes more explicit as the
number of particles is increasing from $k=1$ to $k=5$.

\begin{figure}[ptb]%
\centering
\includegraphics[
height=5.0303cm,
width=7.5981cm
]%
{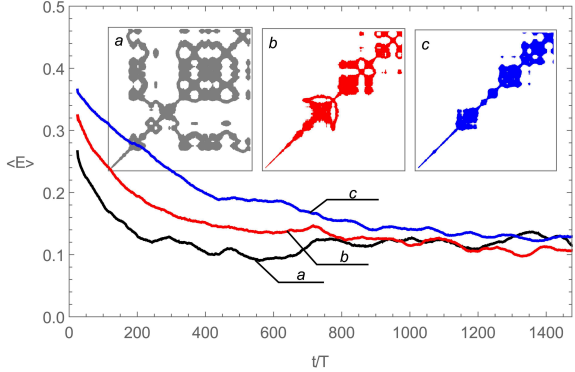}%
\caption{The running 50$T$ average of the container's energy with recurrence
diagrams for different container's shapes: a) $\beta=-0.5$, b) $\beta=-0.2$,
and c) $\beta=0.3$; the horizontal and vertical axes of recurrence diagrams
cover the intervals $0\leq i,j\leq1500$; the number of particles and wall
stiffness parameter are fixed as $k=3$ and $n=4$, respectively. }%
\label{fig9}%
\end{figure}

In addition to the direct visualization of the energy's time histories, we use
a convenient tool of recurrence plots \cite{Eckmann:1987}, \cite{Marwan:2008};
see the upper fragments in Fig.\ref{fig9}. A recurrence plot depicts the
collection of pairs of times at which the trajectory crosses the same
$\varepsilon$ - neighborhood of the system phase space. In particular,
defining the distance in terms of energies, the recurrence (or non-recurrence)
can be characterized by the binary function defined on the two-dimensional
integer grid as%
\begin{equation}
R(i,j)=\left\{
\begin{array}
[c]{ll}%
1\text{,} & \left\vert E_{i}-E_{j}\right\vert \leq\varepsilon\\
0\text{,} & \text{otherwise}%
\end{array}
\right.  \label{Rij}%
\end{equation}
where the energy snapshots $E_{k}=E(kT)$ are taken at every period of the main
oscillator $T=2\pi$.%
%TCIMACRO{\FRAME{ftbpFU}{7.5981cm}{5.0676cm}{0pt}{\Qcb{Container's energy
%during the temporal interval $0\leq t/T\leq1500$ in the case of convex
%scatters ( $\beta=0.3$ ) with the following numbers of particles: a) $k=1$, b)
%$k=3$, and c) $k=5$; the averaging is taken over ensemble of 25 runs over
%random initial positions of particles in the domain \{$-0.01<x(0)<0.01$,
%$-0.01<y(0)<0.01$\}.}}{\Qlb{fig8}}{figure8.eps}%
%width 7.5981cm;  height 5.0676cm;  depth 0pt;  original-width 3.3615in;
%original-height 2.2312in;  cropleft "0";  croptop "1";  cropright "1";
%cropbottom "0";  filename 'Figure8.eps';file-properties "XNPEU";}} }%
%BeginExpansion
%EndExpansion
%TCIMACRO{\FRAME{ftbpFU}{7.5981cm}{5.0303cm}{0pt}{\Qcb{The running 50$T$
%average of the container's energy with recurrence diagrams for different
%container's shapes: a) $\beta=-0.5$, b) $\beta=-0.2$, and c) $\beta=0.3$; the
%horizontal and vertical axes of recurrence diagrams cover the intervals $0\leq
%i,j\leq1500$; the number of particles and wall stiffness parameter are fixed
%as $k=3$ and $n=4$, respectively. }}{\Qlb{fig9}}{figure9.eps}%
%width 7.5981cm;  height 5.0303cm;  depth 0pt;  original-width 6.2483in;
%original-height 4.1156in;  cropleft "0";  croptop "1";  cropright "1";
%cropbottom "0";  filename 'Figure9.eps';file-properties "XNPEU";}} }%
%BeginExpansion
%EndExpansion

In the coordinate plane $i-j$, ones are shown with colors, whereas zeros
remain white; see for instance the corresponding fragments in Fig. \ref{fig9}.
According to the definition (\ref{Rij}), all the recurrence plots are
symmetric with respect to the diagonal $i=j$, on which $\left\vert E_{i}%
-E_{j}\right\vert =0$, and therefore $R(i,j)=1$ for any $i=j$. A non-diagonal
point $(i,j)$ associates the recurrence time as $t_{rec}=T\left\vert
i-j\right\vert $. Note that frequent collisions of small particles against
potential walls produce short-term fluctuations of the energy $E$ generating a
relatively narrow cloud of points near the diagonal $i=j$ with very short
recurrence times. From the standpoint of long-term trends, however, the
meaningful points should distance from the diagonal in such a way that the
longer trend is of interest the longer distance must be considered.

As already mentioned, Fig.\ref{fig8} illustrates long-term temporal behaviors
of the mean energy
%TCIMACRO{\TEXTsymbol{<}}%
%BeginExpansion
$<$%
%EndExpansion
$E$%
%TCIMACRO{\TEXTsymbol{>} }%
%BeginExpansion
$>$
%EndExpansion
at different numbers of particles inside the container. In all the cases, the
initial energy is $E=1/2$. Since all the particles are initially in rest, the
energy of main oscillator
%TCIMACRO{\TEXTsymbol{<}}%
%BeginExpansion
$<$%
%EndExpansion
$E$%
%TCIMACRO{\TEXTsymbol{>} }%
%BeginExpansion
$>$
%EndExpansion
always drops quite abruptly, when the potential wall first collides with the
particle(s). However, further energy behaviors are dictated by the number of
particles and the parameter $\beta$. In particular, the long-term energy decay
is developing more clearly as the number of particles increases. Further, Fig.
\ref{fig9} shows time histories with the corresponding recurrence plots of the
mean energy
%TCIMACRO{\TEXTsymbol{<}}%
%BeginExpansion
$<$%
%EndExpansion
$E$%
%TCIMACRO{\TEXTsymbol{>} }%
%BeginExpansion
$>$
%EndExpansion
at different positive and negative magnitudes of the parameter $\beta$.
Although all of the numbers $\beta$ belong to the NNM instability intervals,
the instability develops in different rates due to different Floquet
multipliers. Note that some recurrence effects are observed in the case of
Buminovich type potential contour; see the case (a) in Fig.\ref{fig9}. This
can be explained more clearly by comparing plots in Fig.\ref{fig9}(a), (b),
and (c) with trajectories of the particle inside potential wells shown in Fig.
4(a), (c), and (f), respectively. Obviously, in the case (a), $\beta=-0.5$,
the trajectory is less chaotic, therefore some recurrence effects can be expected.

\section{Generalizations}

Let us modify the potential function as%

\begin{align}
V  &  =\frac{\gamma}{2n}\left[  \left(  \frac{x}{\alpha+\beta(y^{2}%
-1)}\right)  ^{2n}\right. \nonumber\\
&  \left.  +\left(  \frac{y}{\alpha+\beta(x^{2}-1)}\right)  ^{2n}\right]
\label{V definition2}%
\end{align}

\begin{figure}[ptb]%
\centering
\includegraphics[
height=4.5844cm,
width=8.1012cm
]%
{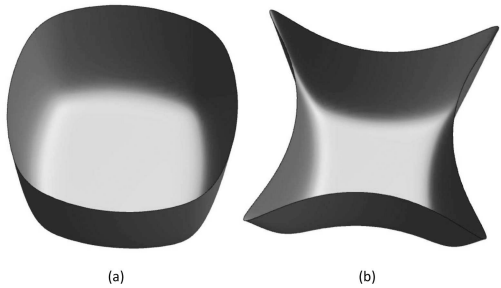}%
\caption{Potential well shapes: a) $\beta=-0.1$- stadium type well, and b)
$\beta=0.1$- convex scattering walls; $\alpha=1.0$, $\gamma=1.0$, $n=6$.}%
\label{fig10}%
\end{figure}

\begin{figure}[ptb]%
\centering
\includegraphics[
height=5.5487cm,
width=8.1012cm
]%
{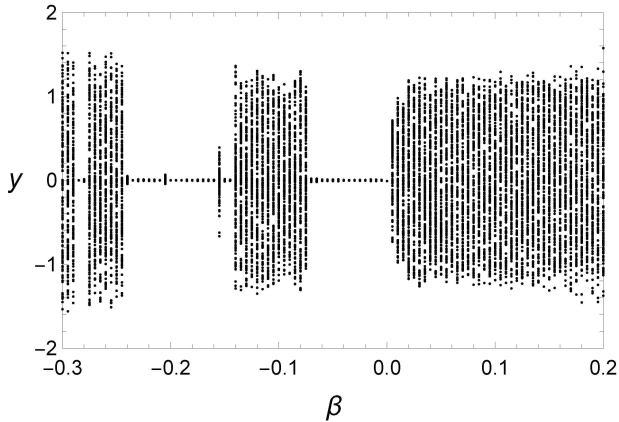}%
\caption{Bifurcation diagram for the case of a single particle, $k=1$, inside
the symmetric potential well (\ref{V definition2}) obtained with parameters:
$\alpha=1.0$, $\gamma=1.0$, $n=6$.}%
\label{fig11}%
\end{figure}

Compared to (\ref{V definition}) the potential well (\ref{V definition}) is
symmetric $x\rightleftarrows y$. For instance, if $\beta>0$ then all the four
pieces of contour's boundary are convex scatters; see Fig.\ref{fig10}.
Qualitatively, the corresponding bifurcation diagram, which is shown in Fig.
\ref{fig11}, is similar to the previous case of flat horizontal boundaries;
compare to Fig. \ref{fig3}. However, there are some numerical differences. For
example, the number $\beta=1.0$ belongs now to the unstable domain. Therefore,
both the potential wells shown in Fig. \ref{fig10} generate the stochastic
dynamics, and therefore the energy absorbing property of inner particles.
%TCIMACRO{\FRAME{ftbpFU}{8.1012cm}{4.5844cm}{0pt}{\Qcb{Potential well shapes:
%a) $\beta=-0.1$- stadium type well, and b) $\beta=0.1$- convex scattering
%walls; $\alpha=1.0$, $\gamma=1.0$, $n=6$.}}{\Qlb{fig10}}{figure10.eps}%
%width 8.1012cm;  height 4.5844cm;  depth 0pt;  original-width 3.1531in;
%original-height 1.772in;  cropleft "0";  croptop "1";  cropright "1";
%cropbottom "0";  filename 'Figure10.eps';file-properties "XNPEU";}} }%
%BeginExpansion
%EndExpansion
%TCIMACRO{\FRAME{ftbpFU}{8.1012cm}{5.5487cm}{0pt}{\Qcb{Bifurcation diagram for
%the case of a single particle, $k=1$, inside the symmetric potential well
%(\ref{V definition2}) obtained with parameters: $\alpha=1.0$, $\gamma=1.0$,
%$n=6$.}}{\Qlb{fig11}}{figure11.eps}{\special{ language "Scientific Word";
%type "GRAPHIC";  maintain-aspect-ratio TRUE;  display "USEDEF";
%valid_file "F";  width 8.1012cm;  height 5.5487cm;  depth 0pt;
%original-width 3.5968in;  original-height 2.4543in;  cropleft "0";
%filename 'Figure11.eps';file-properties "XNPEU";}} }%
%BeginExpansion
%EndExpansion
%TCIMACRO{\FRAME{ftbpFU}{7.6201cm}{5.0808cm}{0pt}{\Qcb{Simulation results in
%the case of five particles, $k=5$, with the potential well parameters:
%$\alpha=1.0$, $\gamma=1.0$, $n=6$, and $\beta=1.0$; the averaging is taken
%over the ensemble of 25 runs and then processed with five periods, 5$T$,
%running average; the initial energy of the container, \TEXTsymbol{<}%
%$E$\TEXTsymbol{>}=0.5 is shown by the dot $t=0$.}}{\Qlb{fig12}}{figure12.eps}%
%width 7.6201cm;  height 5.0808cm;  depth 0pt;  original-width 4.9718in;
%original-height 3.1453in;  cropleft "0";  croptop "1";  cropright "1";
%cropbottom "0";  filename 'Figure12.eps';file-properties "XNPEU";}} }%
%BeginExpansion
\begin{figure}[ptb]%
\centering
\includegraphics[
height=5.0808cm,
width=7.6201cm
]%
{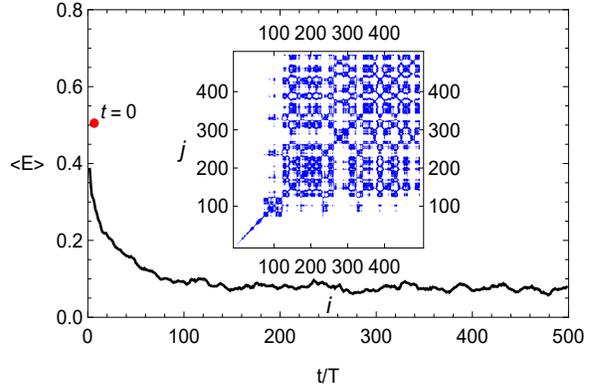}%
\caption{Simulation results in the case of five particles, $k=5$, with the
potential well parameters: $\alpha=1.0$, $\gamma=1.0$, $n=6$, and $\beta=1.0$;
the averaging is taken over the ensemble of 25 runs and then processed with
five periods, 5$T$, running average; the initial energy of the container,
$<$$E$$>$=0.5 is shown by the dot $t=0$.}%
\label{fig12}%
\end{figure}
%EndExpansion

The result of simulation for the case of convex scatters, $\beta=0.1$, is
illustrated in Fig. \ref{fig12}. In particular, it shows that the `statistical
equilibrium' at about
%TCIMACRO{\TEXTsymbol{<}}%
%BeginExpansion
$<$%
%EndExpansion
$E$%
%TCIMACRO{\TEXTsymbol{>} }%
%BeginExpansion
$>$
%EndExpansion
= 0.1 is reached after approximately 100 vibration cycles of the container.
Then the total container energy is fluctuating near the value
%TCIMACRO{\TEXTsymbol{<}}%
%BeginExpansion
$<$%
%EndExpansion
$E$%
%TCIMACRO{\TEXTsymbol{>} }%
%BeginExpansion
$>$
%EndExpansion
= 0.1 with quite small amplitudes; see both the time history and the
recurrence plot in Fig. \ref{fig12}. Therefore particles inside the symmetric
container appear to absorb the energy more effectively compared to the case of
flat boundaries (\ref{V definition}).

\section{Conclusions}

Analyses of the suggested model revealed the existence of effective `energy
sinks' in low-dimensional Hamiltonian systems with potential wells whose
contours, in the rigid-body limit, can take different shapes resembling the
so-called Sinai and Buminovich billiards. We found that the energy flow shows
the one-directional trend during a reasonably long time in the eigen temporal
scale of the system. The analogy with billiards combined with the nonlinear
normal modes stability concept allowed us to determine such parameter
intervals in which light particles inside the potential container possess the
energy absorbing property. On macro-levels, the suggested model can be used
for the design of energy absorbing (harvesting) devices. This can be done by
making container shapes similar to the potential contours revealed in the
present work. Periodic arrays of nano-cells with different types of
`stochastic shapes' can be considered as well in order to design new energy
absorbing materials. In reality, the effect of energy absorption will be
enhanced due to inelastic collisions of particles with container walls. On
molecular levels, different `potential containers' can be created by heavier
particles of crystal latices with arrays of lighter inclusions. For instance,
the repulsive component of Lienard-Jones potential, considered on layers of
cubic latices, can have shapes similar to that shown in Fig.\ref{fig10}(b).

\bibliographystyle{plain}
\bibliography{Bib,BibAuthors,tet}

\end{document}